\documentclass[prb,aps,preprint]{revtex4-1}
\usepackage{graphicx}
\usepackage[pdftex, colorlinks=true, linkcolor=red, citecolor=blue, urlcolor=magenta, pdfstartview=FitH]{hyperref}
\usepackage{color, soul}

\begin{document}

\renewcommand{\thefootnote}{\fnsymbol{footnote}}
\draft

\title{Quantum well photoelastic comb for ultra-high frequency cavity optomechanics}

\author{V. Villafa\~ne$^{1}$, S. Anguiano$^{1}$, A. E. Bruchhausen$^{1}$, G. Rozas$^{1}$, J. Bloch$^{2}$, C. Gomez Carbonell$^{2}$, A. Lema\^itre$^{2}$, and A. Fainstein$^{1, }$\footnote{email:afains@cab.cnea.gov.ar}}
\affiliation{$^1$Centro At\'omico Bariloche \& Instituto Balseiro, C.N.E.A.,8400 S. C. de Bariloche, R. N., Argentina}
\affiliation{$^2$Centre de Nanosciences et de Nanotechnologies, C.N.R.S., Univ. Paris-Sud, Universit\'e Paris-Saclay, C2N Marcoussis, 91460 Marcoussis, France}

\date{\today}

\begin{abstract}

Optomechanical devices operated at their quantum limit open novel perspectives for the ultrasensitive determination of mass and displacement, and also in the broader field of quantum technologies.
The access to higher frequencies implies operation at higher temperatures and stronger immunity to environmental noise. We propose and demonstrate here a new concept of quantum well photoelastic comb for the efficient electrostrictive coupling of light to optomechanical resonances at hundreds of GHz in semiconductor hybrid resonators. A purposely designed ultra-high resolution Raman spectroscopy set-up is exploited to evidence the transfer of spectral weight from the mode at 60 GHz to modes at 190-230 GHz, corresponding to the $8^{th}$ and $10^{th}$ overtone of the fundamental breathing mode of the light-sound cavities.
The coupling to mechanical frequencies two orders of magnitude larger than alternative approaches is attained without reduction of the optomechanical constant $g_0$. The wavelength dependence of the optomechanical coupling further proves the role of resonant photoelastic interaction, highlighting the potentiality to access strong-coupling regimes. The experimental results show that electrostrictive forces allow for the design of devices optimized to selectively couple to specific mechanical modes.  Our proposal opens up exciting opportunities towards the implementation of novel approaches applicable in quantum and ultra-high frequency information technologies.

\end{abstract}

\maketitle


Quantum coherent control of mechanical motion in atomic systems has been exploited since the early
pioneering experiments with trapped ions.~\cite{Ligo,Leibfried,Blatt,Wineland,Cirac} Nano and micromechanical
structures based on condensed matter devices extend these concepts with a great flexibility in design and the possibility to integrate different physical degrees of freedom. In addition, solid state devices allow the access to much higher mechanical frequencies, a critical requirement for quantum operation at higher temperatures and for the implementation of efficient ultrafast quantum information protocols.~\cite{Marquardt}
The possibility to tailor the optical forces by sample design using material dependent electrostrictive forces was theoretically proposed in the context of waveguide optomechanics in Ref.~\onlinecite{Rakich1}.
Here we propose and demonstrate through ultra-high resolution Raman spectroscopy, a quantum well photoelastic comb as a mean for the selective electrostrictive coupling of light to specific mechanical modes, and as a path to cavity optomechanics in the hundreds of GHz range, two orders of magnitude larger than alternative demonstrated technologies.

Most cavity optomechanical devices operate with vibrational frequencies $f_m$ below or in the MHz range, with some designs pushing that limit to a few GHz.~\cite{Marquardt} Attaining higher mechanical frequencies without compromising other operational parameters is of critical relevance for various reasons.  Firstly, to initialize a mechanical oscillator in the ground state at thermal equilibrium, the condition $k_B T/\hbar \Omega_m << 1$ has to be realized (here $\Omega_m=2\pi f_m$). Reaching the ground state (that is, the condition $\overline{n} < 1$, with $\overline{n}$ the average occupation number) is challenging for low-frequency oscillators, with ground-state cooling with conventional cryogenics requiring at least GHz oscillators.  Secondly, higher frequencies are also relevant for displacement measurements.  Quantum fluctuations of optical forces impose a limit on how accurately the position of a free test mass (e.g., a mirror) can be measured.~\cite{Braginsky-Manukin,Braginsky-Khalili,Caves,Jaekel,Pace}  The so-called standard quantum limit determines the minimum possible phonon number,  which defines the ideal performance for continuous position detection, critical e.g. in gravitational wave detectors such as LIGO or VIRGO. The resolved side-band condition $\Omega_m>\kappa$, with $\kappa$ being the optical (photon) dissipation rate is searched for to realize mechanical ground state cooling with light.~\cite{Marquardt}  And thirdly, the $Qf$ product is a direct measure for the degree of decoupling from the thermal environment. Specifically, $Q_m f_m > k_BT/\hbar$ is the condition for neglecting thermal decoherence over one mechanical period. Here the mechanical quality factor is $Q_m = \Omega_m/\Gamma_m$, with $\Gamma_m$ the mechanical (phonon) dissipation rate, $k_B$ is the Boltzman constant and $T$ the system temperature.~\cite{Marquardt}

As mentioned above, the state-of-the-art in high frequency cavity optomechanics has reached the few GHz range.\cite{Ding,Eichenfield,Sun,VanLear}  Reaching higher mechanical frequencies has been hampered so far by the quality limits of top-down nanofabrication techniques such as lithography and etching, as well as the lack of suitable detection methods for the associated fast and minute mechanical motions. Recent work has shown that the so-called ``Extremely High Frequency Range''($\sim 19-95$~GHz) is available using Distributed Bragg Reflector (DBR) GaAs/AlAs semiconductor optomechanical resonators grown by molecular beam epitaxy methods (MBE), with three dimensional optical and mechanical confinement.~\cite{FainsteinPRL2013,AnguianoPRL2017} In these semiconductor materials photons can exert stress through radiation pressure~\cite{Cohadon}, electrostriction (linked to the materials photoelasticity)~\cite{Rakich1,Rakich2}, thermal forces~\cite{MetzgerPRL2008,MetzgerPRB2008,Restrepo}, and so-called optoelectronic forces based on deformation potential interaction involving real photoexcited carriers.~\cite{Okamoto1,Okamoto2,Okamoto3,VillafañeOptoelectronicPRB} In structures based on GaAs/AlAs materials the introduction of quantum wells (QWs) allows an additional degree of freedom to tailor these optical forces. The idea to use QWs for engineering the dynamics of photoexcited carriers operative in optoelectronic forces evidenced in time-resolved experiments with ultrafast pulsed lasers tuned with the absorption gap was reported in Ref.~\onlinecite{VillafañeOptoelectronicPRB}.
It turns out that below the absorption gap photoelastic mediated electrostrictive forces can become the leading contribution if electronic resonances are approached with continuous wave (cw) in hybrid optomechanical resonators containing direct bandgap materials as for example GaAs.~\cite{FainsteinPRL2013,Rozas_Polariton,Baker} In these Brillouin-Raman processes no real excitation of electron-hole pairs, as in Ref.~\onlinecite{VillafañeOptoelectronicPRB}, occurs. We will experimentally show here that this feature can be exploited using highly localized excitonic resonances in quantum wells to define photoelastic combs conceived to strongly, selectively, and efficiently couple through electrostrictive forces confined photon states with specific ultra-high frequency mechanical vibrations of the resonators.

\begin{figure}[!hht]
    \begin{center}
    \includegraphics[trim = 0mm 0mm 0mm 0mm,clip,scale=0.5,angle=0]{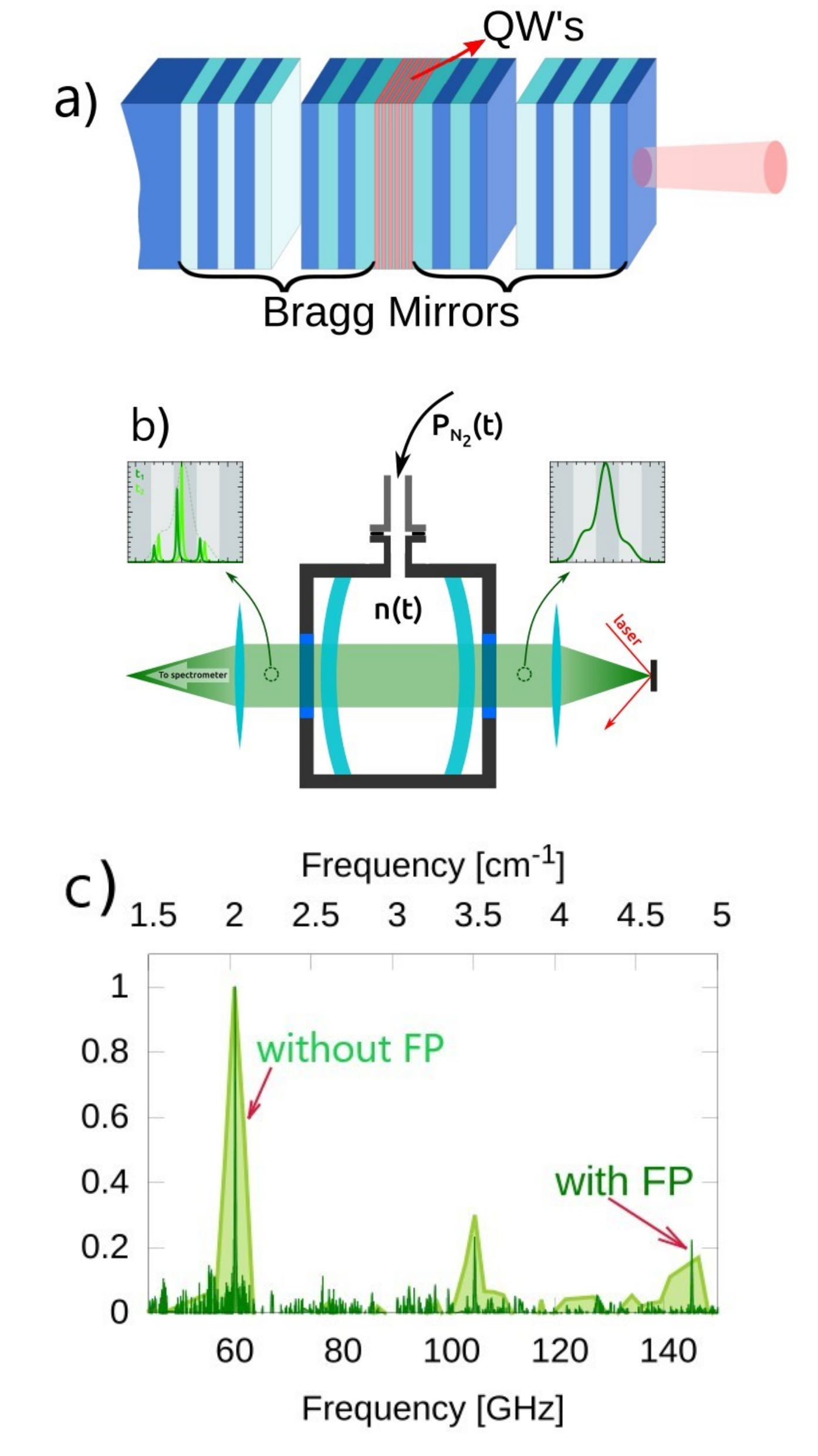}
    \end{center}
    \vspace{-0.8 cm}
\caption{(Color online) .(a) Scheme of the MQW DBR cavity structure. (b) Scheme of the tandem Fabry-Perot (FP) multichanel spectrometer. A FP interferometer with controlled N$_2$ pressure is used to spectrally filter the collected light, prior to its spectral dispersion with a triple spectrometer. A schematic representation of the light spectra relative to the scale of the CCD pixels before and after passage through the FP filter is presented. (c) Example of a real spectrum collected without the FP filter (light green) and with the filter (dark green). \label{Fig1}}
\end{figure}

We consider two {\em planar} microcavity structures, specifically a ``bulk'' GaAs and a multiple quantum well (MQW) resonator. Both cavity structures were grown by molecular beam epitaxy (MBE) on (001)-oriented GaAs substrates. The ``bulk'' GaAs microcavity, included to define a standard to which the QW comb structure will be compared, is made of a $\lambda/2$ uniform GaAs-spacer enclosed by ($\lambda/4,\lambda/4)$ Al$_{0.18}$Ga$_{0.82}$As /AlAs  DBRs, 24 pairs of layers on the bottom, 28 on top, grown on a GaAs substrate.~\cite{Tredicucci,AFainsteinBulkGaAs} As we have demonstrated previously, this structure performs as an optomechanical resonator that simultaneously confines photons and acoustic phonons of the same wavelength.~\cite{FainsteinPRL2013,AnguianoPRL2017} In the MQW microcavity the $\lambda/2$ spacer is constituted by six $145 \mathrm{\AA}$ GaAs QWs separated by $61 \mathrm{\AA}$ AlAs barriers. The DBRs in this case are Al$_{0.10}$Ga$_{0.90}$As /AlAs  multilayers, 27 pairs on the bottom, 23 on top, grown again on a GaAs substrate. A scheme of this structure is displayed in Fig.~\ref{Fig1}(a). Because of the resonant character of photoelastic (electrostrictive) coupling,~\cite{JusserandPRL2015} the associated optical force for photon energies close to the QW exciton resonance will be strongly localised at the QWs, and essentially zero everywhere else.  The fundamental concept behind this design is thus to spatially place the QWs in positions where the strain related to the targeted cavity mechanical modes is maximum. We will term this as a ``quantum well photoelastic comb".

The number of DBR periods in both structures is designed to assure an optical Q-factor $Q \geq 10^4$ (cavity photon lifetime $\tau \sim 5$~ps). Because the contrast of index of refraction and acoustic impedance in the GaAs/AlAs family of materials is coincidently the same,~\cite{FainsteinPRL2013}  this also implies that the nominal {\em mechanical} Q-factors of the high-frequency ($\geq 20$~GHz) mechanical modes will also be in the $Q \geq 10^4$ range. This imposes strong requirements for the spectral resolution and bandwidth of the vibrational spectroscopy used. Spectral noise measurements based on telecommunication technologies as typically used in cavity optomechanics experiments are not applicable in this ultra-high frequency domain. To this aim we have used a purposely developed Raman spectroscopy technique based on a tandem Fabry-Perot triple spectrometer multichannel set-up.~\cite{RozasRSI}

\begin{figure*}[!hht]
    \begin{center}
    \includegraphics[trim = 0mm 40mm 0mm 0mm,clip,scale=0.4,angle=0]{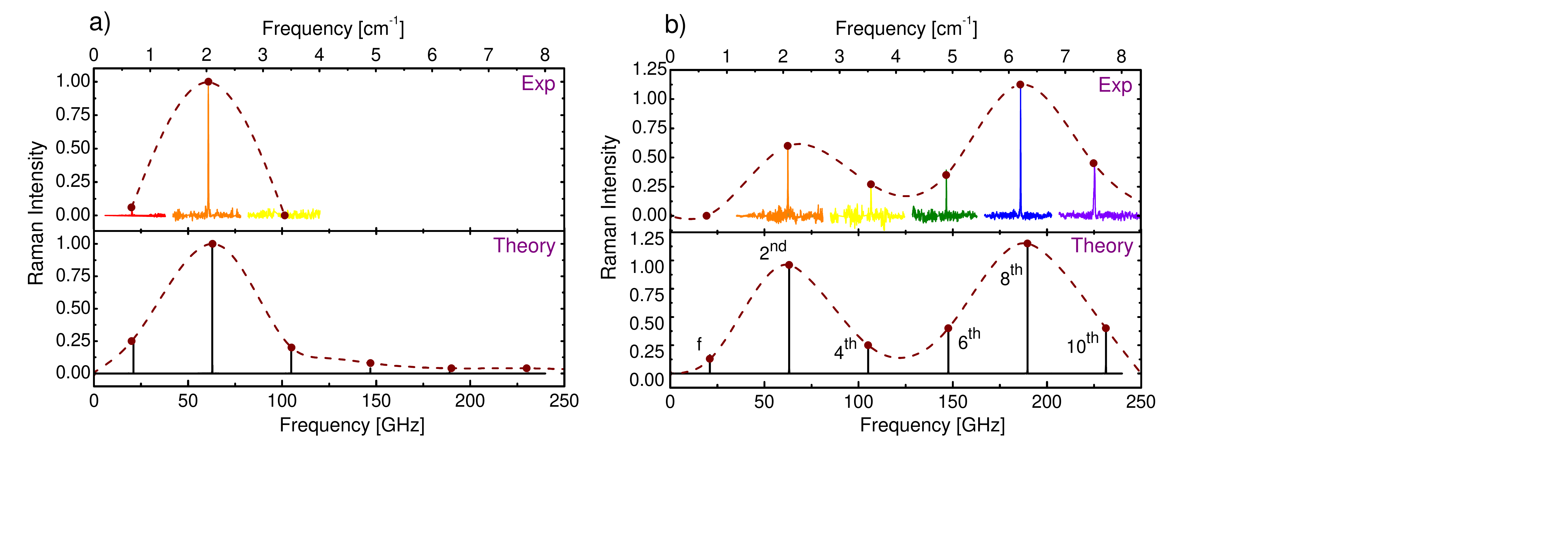}
    \end{center}
    \vspace{-0.8 cm}
\caption{(Color online). Raman spectra for the GaAs (a) and QW photoelastic comb structure (b) at 80~K, corrected by the Bose factor. The top and bottom curves correspond to experiment and model calculations, respectively. Colors in the experimental traces identify separate experiments performed with the double optical resonance (DOR) tuned to selectively enhance the Brillouin-Raman signal corresponding to specific mechanical modes (see text for details). Theoretical calculations are all referred to the intensity of the bulk cavity mode at $\sim 60$~GHz. The experimental intensities for both structures were adjusted to match the magnitude of the corresponding calculated $\sim 60$~GHz mode.  Note the shift of intensity from the 2$^{nd}$ overtone of the fundamental mechanical cavity breathing mode at $\sim 60$~GHz in the GaAs cavity structure, to the 8$^{th}$ and 10$^{th}$ overtones at $\sim 190$  and $\sim 230$~GHz for the MQW device. Dashed curves are guides to the eye. \label{Fig2}}
\end{figure*}

The system is composed of a single-pass Fabry-Perot (FP) interferometer coupled to a T64000 Jobin-Yvon triple spectrometer
operated in additive configuration.~\cite{RozasRSI}  The light to be analyzed is collected from the sample by a lens, filtered through the FP, and then focused by a second lens into the entrance slit of the spectrometer. The FP contains two high-quality ($\lambda$/200)
dielectric mirrors for the near infrared (99$\%$ peak reflectivity centered at 870 nm), which are kept parallel at a fixed distance by three high-quality ($\lambda$/200) cylindrical silica spacers. The mirrors are located in a sealed chamber connected to a pure Nitrogen gas distribution and vacuum system. As the resolution of the spectrometer is better than the FSR of the FP but not enough to resolve the width of its transmission peaks, the acquired spectrum consists of several broad resolution-limited peaks of which the relevant information is their integrated intensity. By repeating this procedure as a function of the gas pressure, we reconstruct the Raman profile with a sub-pixel resolution improved by two orders of magnitude.~\cite{RozasRSI}  The triple spectrometer is equipped with a liquid-N$_2$ cooled charge-coupled device (CCD) multichannel detector which allows for the parallel acquisition of the spectra transmitted through a large set of FP resonances. The excitation is done using a near-infrared Ti:sapphire single-mode Spectra- Physics Matisse TS ring laser, the wavelength of which can be
locked to an external confocal cavity with a precision better than $2 \times × 10^{-6}$ cm$^{-1}$.
With this set-up the resolution of the triple spectrometer was improved from  $\sim 0.5$~cm$^{-1} \sim 15$~GHz to $\sim 3 \times 10^{-3}$~cm$^{-1} \sim 90$~MHz. A scheme of the method is illustrated in Figs.~\ref{Fig1}b-c. Cavity optomechanics spectroscopy was performed using such Fabry-Perot-triple spectrometer tandem in a double optical resonant (DOR) configuration.~\cite{FainsteinPRL1995} In this DOR geometry both the laser and the scattered light are resonant with optical cavity modes of the structure.
For the planar structure studied this can be accomplished by angle tuning and exploiting the in-plane dispersion of the optical cavity modes.~\cite{FainsteinPRL1995}

Figure~\ref{Fig2} contains the main results of this work. Panels (a) and (b) show respectively the mechanical cavity modes determined by ultra-high resolution Raman scattering on the GaAs cavity and the QW photoelastic comb structure. The top spectra correspond to the (Bose-corrected) experiments, while the bottom spectra are the calculations described below. We emphasize that the top curves condense a set of independent experiments, identified with different colors, and performed using different laser incidence angles so that the double optical resonant (DOR) configuration of Brillouin-Raman scattering~\cite{FainsteinPRL1995} is selectively tuned at each and everyone of the observable mechanical cavity modes. All these experiments are then shown together in a single curve. The experiments where performed with the c.w. laser excitation set approximately 10~nm below the bulk GaAs and QW first direct gaps, respectively, at a temperature of 80~K. Theoretical calculations are all referred to the intensity of the bulk cavity mode at $\sim 60$~GHz. Absolute experimental intensities cannot be determined in our experiment, and the optical alignment varies somewhat from one sample to the other. Consequently, the shown intensities for both structures were multiplied by a single constant, the same for all the different peaks of each sample (either bulk or MQW) to better adjust the experimental intensities to the theoretical model. The $\sim 60$~GHz  2$^{nd}$ overtone of the fundamental mechanical cavity breathing mode (at $\sim 20$~GHz) defines the GaAs cavity spectra (with only tiny contributions observable at the frequencies corresponding to the fundamental and 4$^{th}$ overtones). Contrastingly, several modes characterize the MQW structure with a striking shift of spectral weight to the 8$^{th}$ and 10$^{th}$ overtones at $\sim 190$ and $\sim 230$~GHz.

\begin{figure}[!hht]
    \begin{center}
    \includegraphics[trim = 33mm 80mm 0mm 0mm,clip,scale=0.2,angle=0]{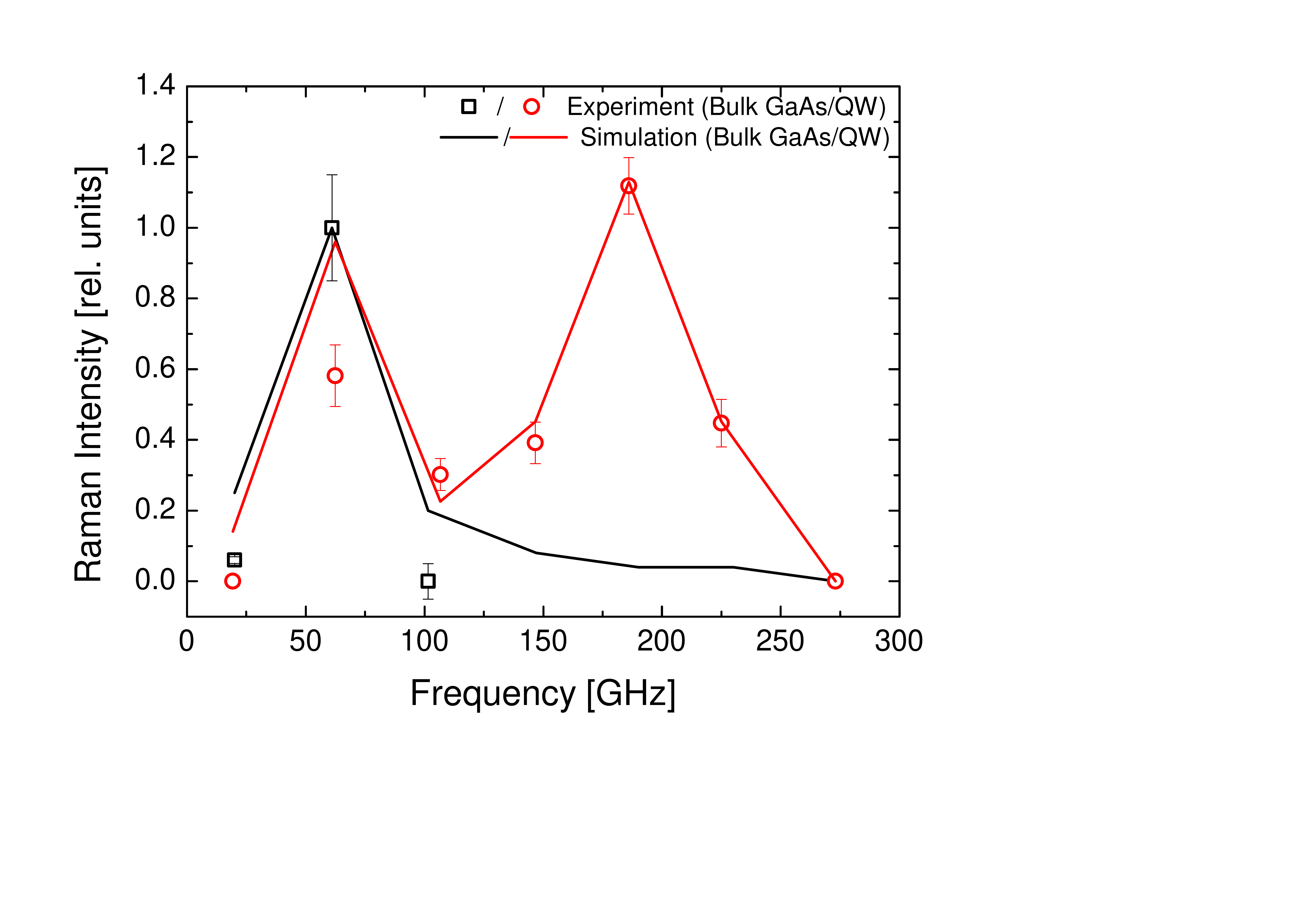}
    \end{center}
    \vspace{-0.8 cm}
\caption{(Color online) . Measured (symbols) and calculated (continuous lines) Raman intensities (integrated peak area) for the bulk and photoelastic comb structures. Theoretical calculations are all referred to the intensity of the bulk cavity mode at $\sim 60$~GHz. A single multiplicative constant was applied to all the different peaks of each sample (either bulk or MQW) to better adjust the experimental intensities to the theoretical model. The Raman intensities have been corrected by the Bose factor (see Eq.~\ref{eq1}).\label{Fig3}}
\end{figure}

The calculated spectra in Fig.~\ref{Fig2} were obtained with a photoelastic model for the Raman cross section that fully takes into account the confined character of optical and mechanical waves and the double optical resonant character of the resonant coupling of laser and Stokes photons to optical cavity modes of the resonator.~\cite{FullTheoryRaman} A direct comparison of the measured and calculated mechanical confined mode intensities (integrated peak area) is provided in Fig.~\ref{Fig3}. In this figure as in Fig.~\ref{Fig2} the measured curves have been corrected by the Bose factor, so that the displayed intensities exclusively represent the optomechanical coupling factor and are not influenced by the thermal phonon population, which is naturally dependent on the involved frequency.
We note that the average phonon number at 80K evaluated using the Bose distribution goes from goes from $\sim 30$ for the  $\sim 60$~GHz mode to $\sim 8$ for the higher frequency $\sim 200$~GHz mode, and can be as low as 1 and 0.1, respectively, at 4~K.  The agreement between experiment and theory evidenced in  Fig.~\ref{Fig3}  is noteworthy, highlighting the applicability of the photoelastic model of optomechanical coupling to describe the involved Raman process.

Qualitatively it is possible to understand the physical origin of the observed mode intensity distribution based on a simplified expression for the Raman cross section $\sigma_R$ ~\cite{LSSV,LSSIX}:
\begin{equation}
\frac{\sigma_R(\Omega_m)}{n(\Omega_m)} \propto \frac{1}{\Omega_m}  \left| \int{p(z)\eta_0(\Omega_m, z)|E(z)|^2}dz \right|^2.
\label{eq1}
\end{equation}
This equation provides $\sigma_R$ corrected by the Bose factor $n(\Omega_m)$, the thermal phonon population. The spatial integral  in Eq.~\ref{eq1} determines the photoelastic contribution to the optomechanical coupling factor $g_0$.~\cite{Baker}  In Eq.~\ref{eq1}, $\Omega_m$ is the phonon angular frequency, $\eta_0$ describes the elastic strain eigenstates, $E(z)$ is the spatially dependent cavity optical mode (which due to the angle DOR tuning is the same for the laser and Stokes scattered fields)~\cite{FainsteinPRL1995}, and $p(z)$ is the material-dependent photoelastic constant. Because of the choice of the laser energy, and its strong resonant behavior, we consider $p(z)$ non-zero only in GaAs (i.e., in the full cavity spacer for the GaAs-cavity, and only in the QWs for the photoelastic comb structure). Due to the Brillouin-Raman scattering configuration, and the orientation of the planar resonators grown along a (001) crystalline axis, coupling only to longitudinal acoustic phonons is allowed, and only one component of the photoelastic tensor is relevant in Eq. 1 ($p_{12}$).~\cite{JusserandPRL2015,LSSV} The situation could be more complex for other crystalline orientations, or for microstructured devices in which z-polarized and in-plane-polarized vibrations get coupled.~\cite{lamberti} We will be interested here in the relative intensity of the vibrational modes, not in their absolute values so that the main physical ingredients are expressed in the functional form of Eq.~\ref{eq1}. This equation essentially reflects the spatial overlap of the strain eigenstates with the cavity confined photon intensity in those regions where their coupling (given by $p$) is non-zero.

\begin{figure}[!hht]
    \begin{center}
    \includegraphics[trim = 0mm 0mm 4mm 0mm,clip,scale=0.375,angle=0]{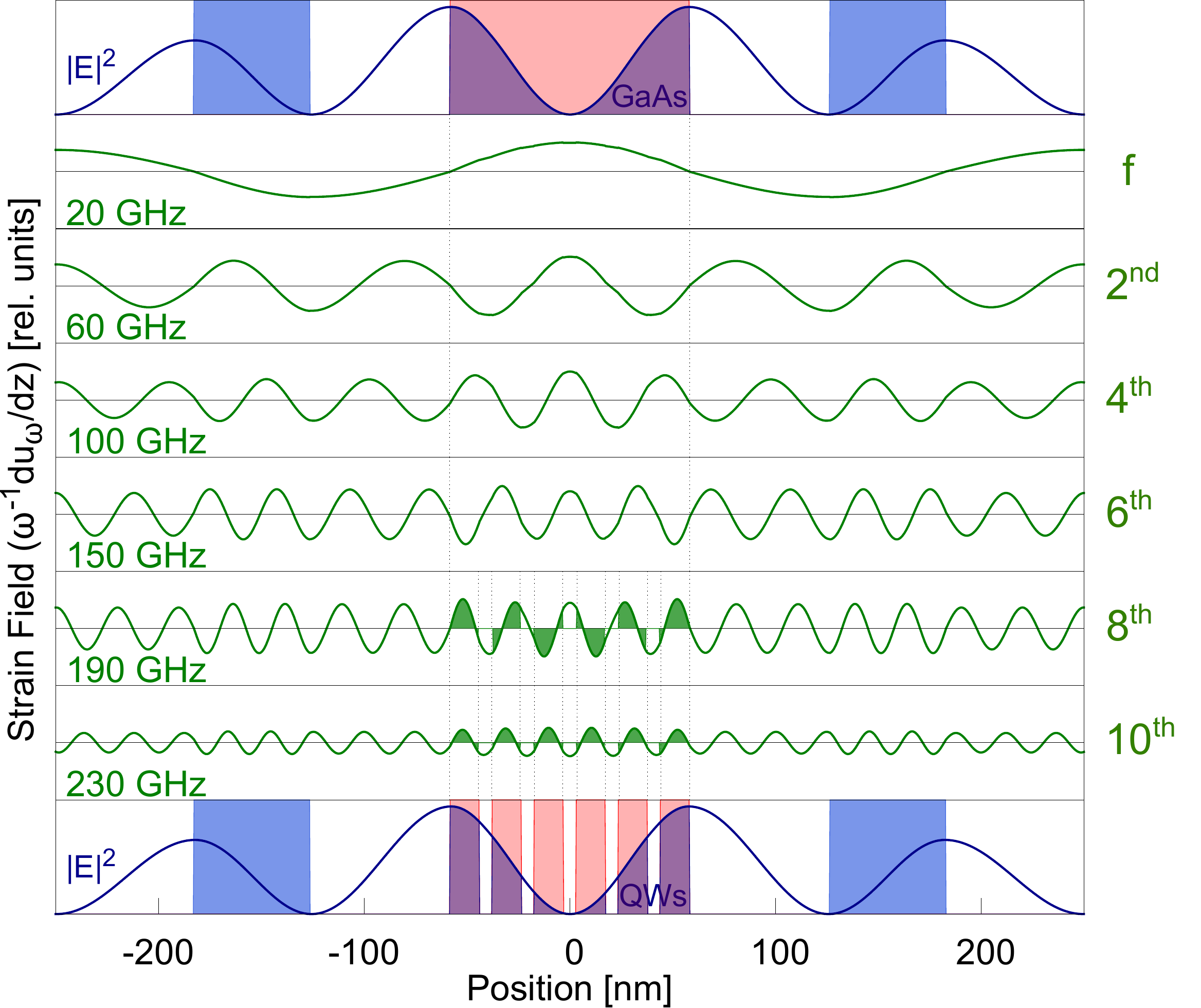}
    \end{center}
    \vspace{-0.3 cm}
\caption{(Color online) . Scheme showing the cavity confined optical and mechanical modes in the central layers of the cavity structures (GaAs cavity on the top, and QW photoelastic comb structure in the bottom). The calculated square of the photon field, and the spatial derivative of the mechanical displacement (i.e., the strain field $\eta$) corresponding to the fundamental breathing mode and its overtones are shown.\label{Fig4}}
\end{figure}

Figure~\ref{Fig4} presents a scheme showing the cavity confined optical and mechanical modes in the central layers of the cavity structures (GaAs cavity on the top, and QW photoelastic comb structure in the bottom). The calculated square of the photon field, and the spatial derivative of the mechanical displacement (i.e., the strain field $\eta$) corresponding to the fundamental breathing mode and its overtones are shown. For the GaAs cavity is clear that the best overlap is obtained for the 60~GHz mode, in agreement with the experiment. Indeed, the 20~GHz fundamental mode has maximum strain where the optical field is zero, while for higher overtones positive and negative strain lobes progressively compensate each other. This latter compensation can be inhibited by positioning the GaAs resonant material {\em only} at the position of strain lobes of the same sign, thus emerging the concept of QW photoelastic comb. For the MQW structure presented here the targeted mode corresponds to the 230~GHz 10$^{th}$ overtone of the fundamental cavity mode. It turns out, however, that also the 190~GHz 8$^{th}$ overtone results strongly amplified because the two central negative lobes of the corresponding strain field overlapping GaAs QWs (i.e., with finite photoelastic interaction) fall where the optical field is almost null.

\begin{figure}[!hht]
    \begin{center}
    \includegraphics[trim = 20mm 50mm 0mm 0mm,clip,scale=0.23,angle=0]{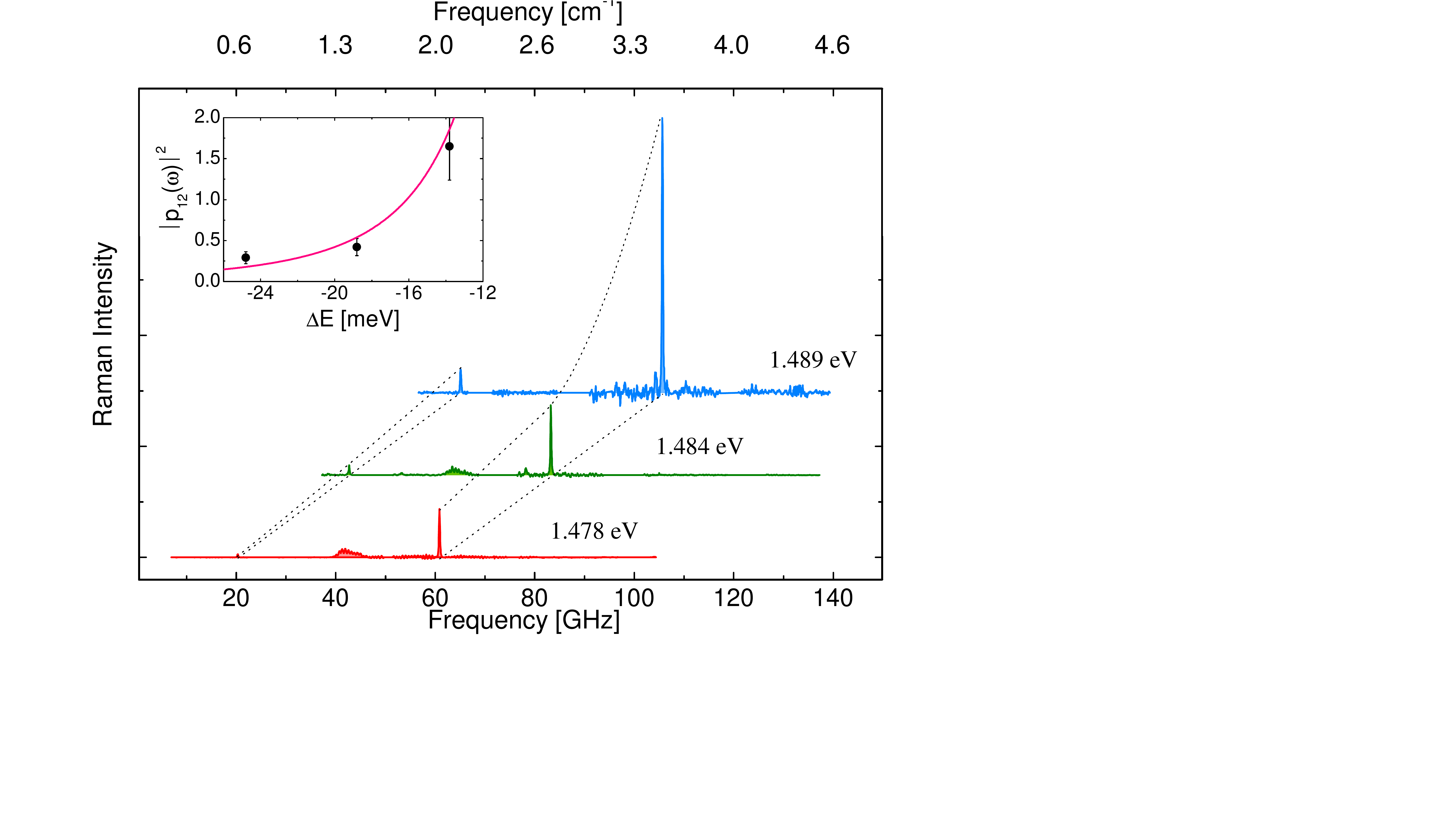}
    \end{center}
    \vspace{-0.8 cm}
\caption{(Color online) . Raman spectra corresponding to the GaAs cavity at 80~K taken at three different laser energies approaching the exciton resonance. The inset displays the measured amplitude of the 60~GHz mode, and the square of the photoelastic constant extracted from the 80~K data of Ref.~\onlinecite{JusserandPRL2015}. To compare with theory the Raman intensities have been scaled with a single multiplicative constant.\label{Fig5}}
\end{figure}

The fact that the optomechanical coupling mechanism is photoelastic (i.e., related to an electrostrictive optical force) is further demonstrated by its resonant behaviour (contained in the laser energy dependence of $p(z)$ in Eq.~\ref{eq1}). Figure~\ref{Fig5} presents three spectra corresponding to the GaAs cavity taken at different laser energies increasing from 1.478~eV up to 1.489~eV, and approaching the 80~K exciton resonance at 1.505eV. A notable enhancement $\sim \times 5$ of phonon amplitude is observed by only blue-shifting 10~meV the laser. The inset in Fig.~\ref{Fig4} shows the amplitude of the 60~GHz mode, compared to the resonant behaviour of the square of the photoelastic constant ($p_{12}$) as determined at 80~K in Ref.~\onlinecite{JusserandPRL2015}. The horizontal scale $\Delta E$ measures the distance to the exciton resonance. Note that only a vertical scale adjustment has been applied to the experimental amplitudes (absolute Raman cross sections are not accessible), evidencing both the strong resonant increase and a qualitative agreement with the expected overall behaviour of the photoelastic resonant enhancement.

In what follows, we explore the magnitude of the single-photon coupling rate $g_0^m$ calculated from the overlap integral of the normalized mode profiles of the optical $\varepsilon(z)$ and the strain field $\eta_0(\Omega_m,z)$\cite{Rakich3}:

\begin{equation}
g_0^m = \epsilon_r q_m \sqrt{\frac{\hbar}{\rho A L_{\mathrm{ac}}\Omega_m}}\frac{\omega}{L_{\mathrm{opt}}}\left| \int{p(z)\eta_0(\Omega_m, z)|\varepsilon(z)|^2}dz \right|,
\label{eq2}
\end{equation}
where $\epsilon_r$, $q_{m}$, $\rho$ and $\omega$ are the medium's relative permittivity, phonon wavevector, mass density and photon angular frequency, respectively. $A$ is the effective area of the optical mode and $L_{\mathrm{opt}}~(L_{\mathrm{ac}})$ is the typical length of the optical (acoustic) interaction. Considering a QW photoelastic comb structure with $Q_{\mathrm{opt}}= 10^4$, $Q_{m}=10^3$, we calculated the optomechanical constant for the $8^{th}$ overtone of angular frequency $\Omega_{m}=2\pi\times190$~GHz. We used $\epsilon_r = 13.7$, $q_m=\Omega_m/v_{\mathrm{ac}}$, $v_{\mathrm{ac}}=4730~\mathrm{m}/\mathrm{s}$, $\rho=5317~\mathrm{kg}/\mathrm{m}^3$, $\omega=2\pi\times365$~THz and $A=\pi(20\mu\mathrm{m}/2)^2$. In order to estimate $L_{\mathrm{opt}}~(L_{\mathrm{ac}})$ we accounted for the fact that in a DBR resonator the optical (acoustic) waves decay exponentially into the mirrors that form the cavity \cite{guille}, thus obtaining $L_{\mathrm{opt}}=L_{\mathrm{ac}}=660$~nm. For the photoelastic constant, we used the value $p_{12}=7$ corresponding to a 6~meV red-shift of the laser with respect of the exciton of the QWs at 80~K.\cite{JusserandPRL2015} Using these parameters, we computed $g_{0}~=~2\pi\times2.2$~MHz. With such high value of optomechanical coupling, the main question is now the threshold power to attain a cooperativity $C=4g_0^2 n_c /(\kappa \Gamma_m) = 1$ ($n_c$ being the intracavity photon number) which leads to zero effective damping and regenerative self-oscillation of the phonon mode. Here $\kappa=\omega/Q_{\mathrm{opt}}=2\pi\times36.5$~GHz and $\Gamma_m=\Omega_m/Q_{m}=2\pi\times190$~MHz are the photon and phonon cavity decay rate, respectively. The intracavity photon number, on the other hand, is obtained from $n_c=2P/(\hbar\omega\kappa)$, assuming that $\kappa=2\kappa_{ext}$. The threshold power results $P\sim$~2mW, consistent with $n_c\approx10^{5}$ and giving an enhanced optomechanical coupling of $g=g_{0}\sqrt{n_{c}}\sim 2\pi\times700$~MHz. Note that the value we used for the photoelastic constant $p_{12}$ is already quite high, due to its resonant character.\cite{JusserandPRL2015} This values could in principle be made even larger by decreasing the temperature and pushing the experiments even closer to the GaAs or QW bandgaps. This resonant character added to the ability to increase further the intracavity photon population highlights the potential of the proposed scheme to access strong-coupling regimes, in which $g>>\kappa$.

In conclusion, we have proposed and demonstrated using double optical resonant Brillouin-Raman experiments a quantum well photoelastic comb to access ultra-high frequency mechanical vibrations of DBR GaAs/AlAs optical microcavities. By placing GaAs quantum wells at the positions of maximum strain and exploiting the resonant character of the photoelastic interaction we have selectively enhanced the optomechanical coupling of high overtones of the fundamental breathing mode of the cavity, showing intense coupling up to 180-230~GHz. We estimated $g_{0}~=~2\pi\times2.2$~MHz for the $8^{th}$ amplified overtone of angular frequency $\Omega_{m}=2\pi\times190$~GHz, giving rise to a threshold power of $\sim 2$~mW to attain a cooperativity equal to unity. The resonant character of the coupling further identifies the photoelastic origin of the interaction, and highlights the potential of the proposed scheme to access strong-coupling regimes. The potentiality of these devices for hybrid designs involving polariton condensates or single quantum dots in addition to the studied quantum wells, inspires intriguing ideas combining different physical degrees of freedom for their application in quantum technologies.

Fruitful discussions with A. Reynoso are acknowledged. This work was partially supported by ANPCyT Grants PICT 2012-1661 and 2013-2047, the International Franco-Argentinean Laboratory LIFAN (CNRS-CONICET) and the French RENATECH network.


\begin{references}

\bibitem{Ligo} B. P. Abbott et al., GW150914: The Advanced LIGO Detectors in the Era of First Discoveries - LIGO Scientific and Virgo Collaborations, Phys. Rev. Lett. {\bf 116}, 131103 (2016).

\bibitem{Leibfried} D. Leibfried, R. Blatt, C. Monroe, and D. Wineland, Quantum dynamics of single trapped ions, Rev. Mod. Phys. {\bf 75}, 281 (2003).

\bibitem{Blatt} R. Blatt, and D. Wineland, 2008, Entangled states of trapped atomic ions, Nature (London) {\bf 453}, 1008 (2008).

\bibitem{Wineland} D. J. Wineland, Nobel Lecture: Superposition, entanglement, and raising Schrödinger?s cat, Rev. Mod. Phys. {\bf 85}, 1103 (2013).

\bibitem{Cirac} J. Cirac,  and P. Zoller, Quantum Computations with Cold Trapped Ions, Phys. Rev. Lett. {\bf 74}, 4091 (1995).

\bibitem{Marquardt} M. Aspelmeyer, T. J. Kippenberg, and F. Marquardt, "Cavity optomechanics," Rev. Mod. Phys. {\bf 86}, 1391 (2014).

\bibitem{Rakich1} P. T. Rakich, P. Davids, and Z. Wang,  Tailoring optical forces in waveguides through radiation pressure and electrostrictive forces, Opt. Express {\bf 18}, 14439– (2010).

\bibitem{Braginsky-Manukin} V. B. Braginsky,  and A. B. Manukin, {\it Measurement of weak forces in Physics experiments} (University of Chicago Press, Chicago, 1977).

\bibitem{Braginsky-Khalili}  V. B. Braginsky, and F. Y. A. Khalili, {\it Quantum Measurements} (Cambridge University Press, Cambridge, England, 1995).

\bibitem{Caves} C. M. Caves, Quantum-Mechanical Radiation-Pressure Fluctuations in an Interferometer, Phys. Rev. Lett. {\bf 45}, 75. (1980).

\bibitem{Jaekel} M. Jaekel, and S. Reyaud, Casimir force between partially transmitting mirrors, J. Phys. I (France) {\bf 1}, 1395 (1991).

\bibitem{Pace} A. Pace, M. Collett, and D. Walls, Quantum limits in interferometric detection of gravitational radiation, Phys. Rev. A {\bf 47}, 3173 (1993).

\bibitem{Ding} L. Ding, C. Baker, P. Senellart, A. Lemaitre, S. Ducci, G. Leo, and I Favero, High Frequency GaAs Nano-Optomechanical Disk Resonator, Phys. Rev. Lett. {\bf 105}, 263903 (2010).

\bibitem{Eichenfield} M. Eichenfield, J. Chan, R. M. Camacho, K. J. Vahala, and O. Painter, Optomechanical crystals, Nature (London) 462, 78 (2009).

\bibitem{Sun} X. Sun, X. Zhang, and H. X. Tang, High-Q silicon optomechanical microdisk resonators at gigahertz frequencies, Appl. Phys. Lett. 100, 173116 (2012).

\bibitem{VanLear} R. Van Laer, B. Kuyken, D. Van Thourhout, and R. Baets, Interaction between light and highly confined hypersound in a silicon photonic nanowire, Nat. Photonics 9, 199 (2015).

\bibitem{FainsteinPRL2013} A. Fainstein, N. D. Lanzillotti-Kimura, B. Jusserand, and B. Perrin, Strong Optical-Mechanical Coupling in a Vertical GaAs/AlAs Microcavity for Subterahertz Phonons and Near-Infrared Light, Phys. Rev. Lett. {\bf 110}, 037403 (2013).

\bibitem{AnguianoPRL2017} S. Anguiano, A. E. Bruchhausen, B. Jusserand, I. Favero, F. R. Lamberti, L. Lanco, I. Sagnes, A. Lema\^itre, N. D. Lanzillotti-Kimura,  P. Senellart, and A. Fainstein, Micropillar Resonators for Optomechanics in the Extremely High 19-95-GHz Frequency Range, Phys. Rev. Lett. {\bf 118}, 263901 (2017).

\bibitem{Cohadon} P. F. Cohadon, A. Heidmann, and M. Pinard, Cooling of a Mirror by Radiation Pressure, Phys. Rev. Lett. {\bf 83}, 3174 (1999).

\bibitem{Rakich2} P. T. Rakich, C. Reinke, R. Camacho, P. Davids, and Z. Wang, Giant Enhancement of Stimulated Brillouin Scattering in the Subwavelength Limit, Phys. Rev. X {\bf 2}, 011008 (2012).

\bibitem{MetzgerPRL2008} C. Metzger, M. Ludwig, C. Neuenhahn, A. Ortlieb, I. Favero, K. Karrai, and F. Marquard, Self-Induced Oscillations in an Optomechanical System Driven by Bolometric Backaction, Phys. Rev. Lett. {\bf 101}, 133903 (2008).

\bibitem{MetzgerPRB2008} C. Metzger, I.Favero, A. Ortlieb, and K. Karrai, Optical self cooling of a deformable Fabry-Perot cavity in the classical limit,  Phys. Rev. B {\bf 78}, 035309 (2008).

\bibitem{Restrepo} J. Restrepo, J. Gabelli, C. Ciuti, and I. Favero, Classical and quantum theory of photothermal cavity cooling of a mechanical oscillator, Comptes Rendus Physique
{\bf 12}, 860 (2011).

\bibitem{Okamoto1} Hajime Okamoto, Daisuke Ito, Koji Onomitsu, Haruki Sanada, Hideki Gotoh, Tetsuomi Sogawa, and Hiroshi Yamaguchi, "Vibration Amplification, Damping, and Self- Oscillations in Micromechanical Resonators Induced by Optomechanical Coupling through Carrier Excitation," Phys. Rev. Lett. {\bf 106}, 036801 (2011).

\bibitem{Okamoto2} Hajime Okamoto, TakayukiWatanabe, Ryuichi Ohta, Koji Onomitsu, Hideki Gotoh, Tetsuomi Sogawa, and Hiroshi Yamaguchi, "Cavity-less on-chip optomechanics using excitonic transitions in semiconductor heterostructures," Nature
Comm. {\bf 6}, 8478 (2015).

\bibitem{Okamoto3} Hajime Okamoto, Daisuke Ito, Koji Onomitsu, Tetsuomi Sogawa, and Hiroshi Yamaguchi, "Controlling Quality Factor in Micromechanical Resonators by Carrier Excitation," Applied Physics Express {\bf 2} (2009) 035001.

\bibitem{VillafañeOptoelectronicPRB} V. Villafa\~ne, P. Sesin, P. Soubelet, S. Anguiano, A. E. Bruchhausen, G. Rozas, C. Gomez Carbonell, A. Lema\^itre, and A. Fainstein, Optoelectronic forces with quantum wells for cavity optomechanics in GaAs/AlAs semiconductor microcavities Phys. Rev. B {\bf 97}, 195306 (2018).

\bibitem{Rozas_Polariton} G. Rozas, A. E. Bruchhausen, A. Fainstein, B. Jusserand, and A. Lema\^itre, Polariton path to fully resonant dispersive coupling in optomechanical resonators, Phys. Rev. B {\bf 90}, 201302(R) (2014).

\bibitem{Baker} C. Baker, W. Hease, Dac-Trung Nguyen, A. Andronico, S. Ducci, G. Leo, and I. Favero, Photoelastic coupling in gallium arsenide optomechanical disk resonators, Optics Express {\bf 22}, 14072 (2014).

\bibitem{Tredicucci} A. Tredicucci, Y. Chen, V. Pellegrini, M Bo¨rger, L. Sorba, F. Beltram, and F. Bassani, Controlled Exciton-Photon Interaction in Semiconductor Bulk Microcavities, Phys. Rev. Lett. {\bf 75}, 3906 (1995).

\bibitem{AFainsteinBulkGaAs} A. Fainstein, B. Jusserand, P. Senellart, J. Bloch, V. Thierry-Mieg, and R. Planel, Center-of-mass quantized exciton polariton states in bulk-GaAs microcavities, Phys. Rev. B {\bf 62}, 8199 (2000).

\bibitem{JusserandPRL2015} B. Jusserand, A.N. Poddubny, A.V. Poshakinskiy, A. Fainstein, and A. Lema\^itre, Polariton Resonances for Ultrastrong Coupling Cavity Optomechanics in GaAs/AlAs Multiple Quantum Wells, Phys. Rev. Lett. {\bf 115}, 267402 (2015).

\bibitem{RozasRSI} G. Rozas, B. Jusserand, and A. Fainstein, Fabry-Pérot-multichannel spectrometer tandem for ultra-high resolution Raman spectroscopy, Review of Scientific Instruments {\bf 85}, 013103 (2014).

\bibitem{FainsteinPRL1995}  A. Fainstein, B. Jusserand, and V. Thierry-Mieg, Raman Scattering Enhancement by Optical Confinement in a Semiconductor Planar Microcavity, Rev. Lett. {\bf 75}, 3764 (1995).

\bibitem{FullTheoryRaman} A. E. Bruchhausen, G. Rozas, and A. Fainstein, Full model for acoustic phonon Raman spectra in multilayer planar optomechanical resonators, to be published.

\bibitem{LSSV} B. Jusserand and M. Cardona. In M. Cardona and G. Guntherodt, editors, Topics in Applied Physics, Light Scattering in Solids V, volume 66, chapter 3, pages 49-152. Springer-Verlag, Berlin Heildelberg New York, 1 edition, 1989.

\bibitem{LSSIX} A. Fainstein and B. Jusserand. In M. Cardona and R. Merlin, editors, Topics in Applied Physics, Light Scattering in Solids IX, volume 108, chapter 2, pages 17-110. Springer-Verlag, Berlin Heildelberg New York, 1 edition, 2007.

\bibitem{lamberti} F. R. Lamberti, Q. Yao, L. Lanco, D. T. Nguyen, M. Esmann, A. Fainstein, P. Sesin, S. Anguiano, V. Villafañe, A. Bruchhausen, P. Senellart, I. Favero, and N. D. Lanzillotti-Kimura, Optomechanical properties of GaAs/AlAs micropillar resonators operating in the 18 GHz range, Opt. Express {\bf25}, 24437 (2017).

\bibitem{Rakich3} P. Kharel, G. I. Harris, E. A. Kittlaus, W. H. Renninger, N. T. Otterstrom, J. G. E. Harris, and P. T. Rakich, High-frequency cavity optomechanics using bulk acoustic phonons, arXiv preprint arXiv:1809.04020 (2018).

\bibitem{guille} D. I. Babic y S. W. Corzine, Analytic expressions for the reflection delay, penetration depth, and absorbance of quarter-wave dielectric mirrors, Institute of Electrical and Electronics Engineers Journal of Quantum Electronics {\bf28}, 514 (1992).



\end{references}

\end{document}